\documentclass[useAMS,usenatbib,usegraphicx]{mn2e}
\usepackage{}

\title{Fast Spectral Variability from Cygnus~X-1}
\author[Y. X. Wu, T. M. Belloni and L. Stella]{Y. X. Wu$^{1,2}$\thanks{E-mail:
wuyx@mails.thu.edu.cn; tomaso.belloni@brera.inaf.it}, T. M. Belloni$^{2}$ and L. Stella$^{3}$\\
$^{1}$Department of Engineering
Physics and Center for Astrophysics, Tsinghua University, Beijing 100084,
China\\
$^{2}$INAF-Osservatorio Astronomico di Brera, Via Bianchi
46, I-23807 Merate, Italy\\
$^{3}$INAF-Osservatorio
Astronomico di Roma, via Frascati 33, I-00040 Monteporzio Catone,
Italy
}
\begin{document}

\date{}

\pagerange{\pageref{firstpage}--\pageref{lastpage}} \pubyear{}

\maketitle

\label{firstpage}

\begin{abstract}
We have developed an algorithm that, starting from the observed properties
of the X-ray spectrum and fast variability of an X-ray binary allows
the production of synthetic data reproducing observables such as power density spectra and time lags, as well as their energy dependence. This allows to reconstruct the variability of parameters of the energy spectrum and to
reduce substantially the effects of Poisson noise, allowing to study fast spectral variations. We have applied the algorithm to Rossi X-ray Timing Explorer data of the black-hole binary Cygnus~X-1, fitting the energy spectrum with a simplified power law model. We recovered the distribution of the power law spectral indices on time-scales as low as 62~ms as being limited between 1.6 and 1.8. The index is positively correlated with the flux even on such time-scales.
\end{abstract}


\section{INTRODUCTION}
Black hole binaries (BHB) exhibit considerable X-ray variability on a wide
range of time-scales. The study of X-ray fast time
variability has become an important astrophysical research tool that
helps us gain better insight into the physical process at work near
the black hole \citep[see the recent review of][]{vdK04}.
For instance, the dynamic time-scale for the motion within a few
Schwarzschild radii of a $10~M_{\odot}$ black hole is at the order of
milliseconds. Further considering that most of the gravitational
energy of accretion matter is released in the inner area of a few
Schwarzschild radii, the variability at short time-scales can be
used to probe the accretion-flow dynamics and geometries within the
strong-field region.

Time variability can be studied in the time domain or in the
frequency domain. The latter is based on the Fourier transform (FT) and usually is based upon two basic techniques, the Power Density Spectrum (PDS) and the time lag spectrum. The square of Fourier transform amplitudes as a function of Fourier frequency constitutes the PDS, which provides the estimate of variance at different frequencies. The time lag spectrum is obtained from the phase lag, i.e. the phase angle difference between the Fourier vectors at different energy channels. In practice, the PDS and the lag spectra are usually averaged over many segments of observation and
frequencies in order to increase the statistical significance. The Fourier
transform is reversible: the time series can be reconstructed from its Fourier transform by means of Inverse Fourier transform (IFT).
On the contrary, the PDS is not reversible, since the phase information in the FT is lost.  In principle there is an infinite variety of different signals that will yield the same PDS.

The fast variability observed from BHB is of stochastic nature and
as such cannot be modeled directly. In other words, it is not
possible to reproduce the exact observed variations. The aim of
time-series analysis is to characterize the average properties that
give rise to the fluctuations, under the assumption that the process
is stationary. A successful model should reproduce the PDS and the
lag spectrum, as well as other statistical properties of the signal
\citep[see, e.g.][]{Utt05}. A conventional model describing the
temporal fluctuation is the shot-noise model \citep{Tel72,Neg94}. It
has become clear, however, that in this framework complex shot
profiles or distributions of shot durations and amplitudes have to be
assumed to model the variability of BHBs \citep[e.g.][]{Miy88,BH90,Loc91}.
An alternative way is to apply Linear State Space Models (LSSMs)
which are based on stochastic processes, or autoregressive (AR)
processes to describe the temporal variability \citep{KT97,Pot98}.
\citet{Utt05} use a non-linear model to explain the lognormal flux
distribution and rms-flux relation. All these models are
phenomenological; based on the PDS alone they try to reproduce the observed properties through a mathematical model, which can provide
constraints on physical models. On the other hand, \citet{AU06} attempted a more physically-constrained generating process to model all the spectral-timing properties simultaneously.

The usual course of action is to extract information in the
frequency domain, such as a PDS, from the time series. However,
sometimes we need to do the opposite: to reconstruct the time
series from the PDS. The simulation of random time series with arbitrary PDS has a long and established history \citep[see e.g.][]{Dav81,LM82} in the field of digital signal processing. There are also papers on simulating time series with specific marginal distribution, e.g. lognormal \citep{Joh94}.
In astrophysical research the reasons and benefits to perform such a reconstruction are various.
Sometimes it provides a more direct tool to judge the models or simulation
methods \citep[for an example, see][]{TK95}. Also it can be used to
estimate the error bars by Monte Carlo simulation
\citep[e.g.][]{Don92}. Since the PDS does not contain the phase
information, in order to do so one must assign values to the phases
as the reconstruction based on the PDS alone is not unique. One easy way to generate data that reproduce a given PDS is to choose the Fourier amplitude according to the PDS and assign random phases between $[0,2\pi]$ \citep{Don92}. Based on the theory of linear stochastic process and the fact that the PDS itself follows a chi-square distribution, \citet{TK95} proposed a algorithm practically identical to \citet{DH87}---to produce the whole variety of possible
non-deterministic linear time series from the PDS by randomizing
both the phases and amplitudes. Some authors use the energy-resolved PSDs instead of the total PSD in the method of \citet{TK95}, and shift the phase to yield light curves with desired lag function between them \citep[e.g., ][]{Zog10}. Recently, the non-linear behavior of observed light curves was studied by \citet{Utt05}. They suggested that an additional exponential transform needs to be applied to the time series created with the method of \citet{TK95}.

BHBs are known to exhibit X-ray spectral evolution on short time-scales.
This evolution is reflected in the presence of lags between the light curves at different energy ranges and asymmetries of the cross-correlation function between them, as well as fast variations of the corresponding
hardness ratio. However, the conventional spectral models applied to
these systems are designed to fit the energy spectrum averaged over,
usually, several thousand seconds. Because of limited statistics, it
is not possible to follow the energy spectrum over the short time-scales corresponding to the observed fast variability.

In this work we propose a new technique that simulates a time series
starting from the actual intensity measurements.
Specifically, we simulate the light curves in different
energy bins reproducing all the properties observed in the real data:
the average energy spectrum, the PDS as a function of energy and the
the frequency-dependent lag spectra between different
energy bands. In the simulation, no Poisson noise is of course
introduced. With the simulated ``clean" light curves in different
energy bins, we can study the variations of the energy spectrum on
short time-scales. Our work is improved (or different) in three aspects when compared with the papers mentioned above. For the first, besides the PDS, we make use of other measurements, including energy spectra and lag spectra as input to the simulation. Secondly, we require that the simulated time series should reproduce almost all the timing and spectral properties. At last we explore the possible application of the method, including filtering Poisson noise and data extrapolation. Our work, based on those by \citet{TK95} and \citet{Utt05}, can be seen as a continuation of them. In contrast to \citet{AU06}, our work is model-independent and aims at developing an algorithm that recovers the time series preserving the information and filtering the noise.

The article is organized as follows: the
preparatory works of data analysis and parameter estimation are
introduced in section~\ref{sec_pre}, including the non-linearity
study (section~\ref{sec_nl}), energy dependence of PDS
(section~\ref{sec_pdsen}), lag spectra (section~\ref{sec_lagen}),
coherence function (section~\ref{sec_coh})
and distribution of the phases (section~\ref{sec_phas}). The
important results are presented in section~\ref{sec_result}. In
section~\ref{sec_alg} the algorithm is defined by steps. The
simulated light curve is compared with the observed one
(section~\ref{sec_test}) and the issue of Poisson noise subtraction
is discussed (section~\ref{sec_subnoise}). As one important
application, the energy spectra on short time-scales (dynamic energy
spectra) are derived and the variation of spectral shape is
investigated in section~\ref{sec_des}. Another possible application,
the simulation of data with better time and energy resolution than
observations can be measured directly from the observations,
is discussed in section~\ref{sec_extint}. The issues
associated with phase and noise are discussed in
section~\ref{sec_posnois}. The conclusions are presented in
section~\ref{sec_con}.

\section{DATA ANALYSIS AND PARAMETER ESTIMATION}\label{sec_pre}

The central idea of the algorithm is to synthesize light curves that
reproduce all observed properties both in the time and frequency domain. These are:

\begin{itemize}
   \item the average count rate in different energy bands both integrated (spectrum)
   and as a function of time (light curves)
   \item the relation between root mean square variability and count rate in the light curves at
   different energies (rms-flux relation)
   \item the shape and normalization of the PDS as a function of energy
   \item the phase/time-lag spectrum as a function of energy
\end{itemize}

Once these properties are extracted from the data, synthetic data
curve can be constructed to reproduce the same results.

A preliminary step is to obtain the above quantities with good accuracy for a black-hole binary. We used a single observation of Cygnus X-1 from the Proportional Counter Array (PCA) on board the Rossi X-ray Timing Explorer ({\it RXTE}) (ObsID 10238-01-05-000). Cyg~X-1 is the first-discovered black hole binary and has been studied for several decades. Its brightness and persistence make it a perfect target for X-ray timing research. The observation was carried out in March, 1996, when Cyg~X-1 was in its hard spectral state, the most common for the source
\citep[e.g.][]{Wil06}. The data configuration used here is the
generic PCA binned mode B\_16ms\_64M\_0\_249, which provides high
resolution in both energy (64 channels over the full 2-60~keV PCA
band) and time (16~msec). This is the main reason why we selected
this particular observation, which also has the advantage of having been made with all five proportional counter units of the PCA, increasing the source count rate. In order to obtain sufficient statistics for the analysis, we rebinned the data into eight energy bins between 0.14--25~keV. In this way in each energy bin the mean count rate is above $\sim1000$~counts~s$^{-1}$. Notice that the contribution of
background photons to each of these eight bins is minor compared to the source counts.

For the analysis, we used custom-made software written in the IDL environment. For each of the eight energy bins, we extracted a Fourier Spectrum from data stretches of length 128~s, up to a Nyquist frequency of 32~Hz. These Fourier spectra were averaged for constructing the PDS \citep[normalized to the squared fractional rms, see][]{BH90} and the phase-lag spectra. The Poisson contribution was subtracted by using RXTE recipes \citep[see][]{Zha95}.

\subsection{Rms-Flux relation}\label{sec_nl}

\begin{figure*}
\includegraphics[width=0.7\textwidth]{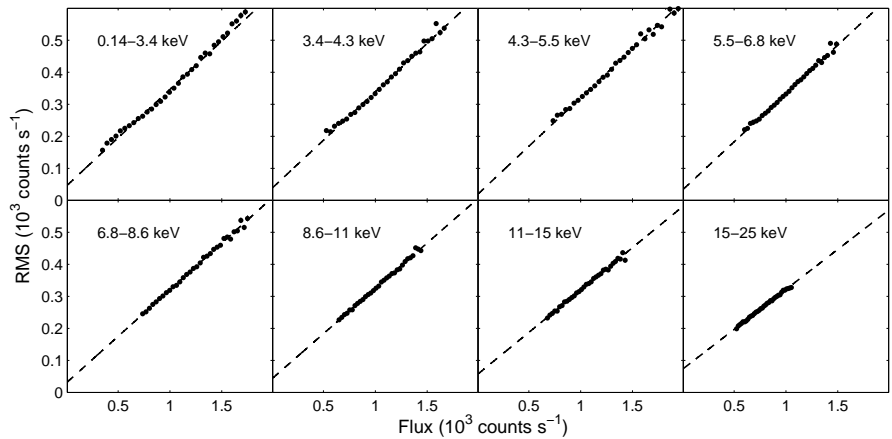}
  \caption{The rms-flux relations for 8 energy bins.
They are produced by binning the 1-32~Hz rms measured in 1~s segments
into flux bins. The dashed line is the best-fitting linear
model for each plot. \label{fig_rmsflux}}
\end{figure*}

\begin{figure*}
\includegraphics[width=0.7\textwidth]{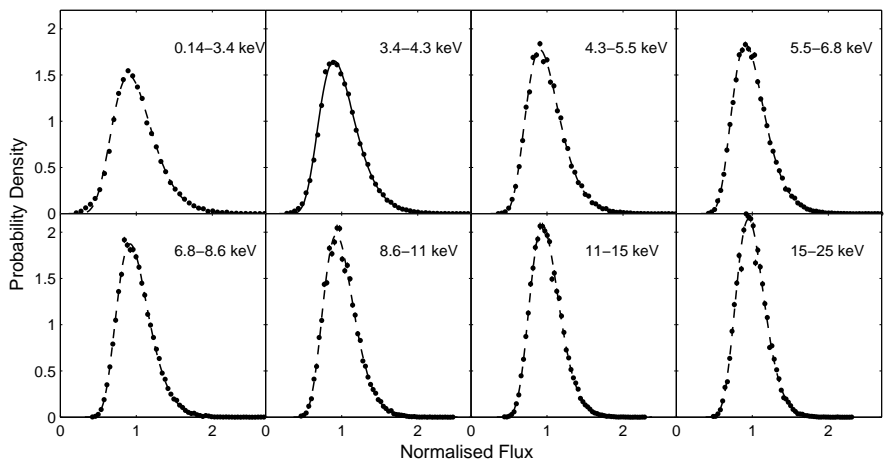}
  \caption{The flux distribution expressed as a
probability density, i.e. the data points per flux bin
normalized by flux bin-width and by the total number of data points.
The flux is calculated over 0.25~s time bins. The dashed line is
the best-fitting lognormal model for each plot.
\label{fig_fluxhis}}
\end{figure*}

\citet{Utt05} showed that the rms-flux relation, the non-linear
behavior and the lognormal flux distribution observed in the hard state of BHB represent three
different aspects of the same underlying process. This non-linearity
can be reproduced as an exponential of a linear light curve. As the
plan is to simulate the source light curve in different energy
bands, we must first check that this non-linearity holds also for
separate energy band. We then produced the rms-flux relation and the
flux (count rate) distribution for each of our eight energy bins (see
Fig.~\ref{fig_rmsflux} and Fig.~\ref{fig_fluxhis}).

The rms in Fig.~\ref{fig_rmsflux} was measured by integrating the PDS of the eight bins over the 1--32~Hz frequency interval
for 1~s segments.  Its relation with the count rate, also in 1-s bins, is consistent with linearity for all bins. This linear relation holds also when the length of the light curve segments and the frequency range for the integration are changed. At the same time, the flux distribution (with a time bin of 0.25~s in Fig.~\ref{fig_fluxhis}) fits a lognormal model.
Further subdividing the energy range into narrower bins we did not find significant deviations.

\begin{figure}
\includegraphics[width=84mm]{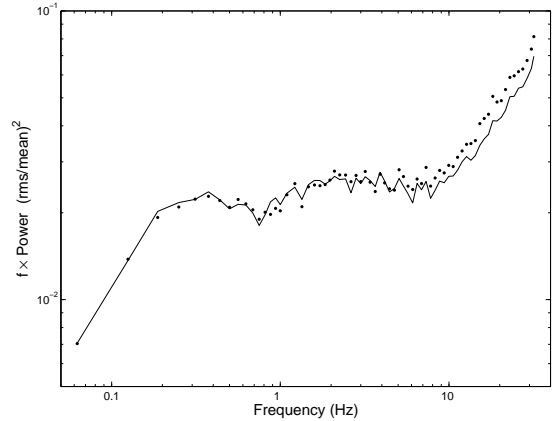}
  \caption{ The PDS of the raw data (solid line) and the logarithmically transformed data (dot). The latter is renormalized to be directly compared with the former. The Poisson noise is not subtracted from the PDS.
\label{fig_pdslog}}
\end{figure}

As shown by \citet{Utt05}, for typical observed light curves with
the PDS dominated by broad components and a fractional rms of
20--40 per cent, the distorting effect of the exponential transformation on the shape of the PDS is relatively small. A quantitative analysis
(not presented here) suggests that for Lorentzian-shaped PDS
components with fractional rms smaller than 50 per cent, the distortion is not serious if the quality factor \footnote{The quality factor is the ratio between the Lorentzian centroid frequency and the full width at half maximum (FWHM).} $Q\la 2$. We tested the distortion on the PDS caused by exponential transformation of data in the time domain: we calculated the PDS from the logarithm of the real data and compared it with the original one (Fig.~\ref{fig_pdslog}). The PDS shapes are almost unchanged between the raw data and the logarithmically transformed data.
Therefore in our simulation below, it is justified to use the observed PDS as the PDS of the input linear light curve without the need to correct the distortion effect of the exponential transformation. However, it is
still important to apply the appropriate correction to the normalization of the input PDS in order to return the desired variance in the output light curve, because the exponential transformation will cause the increase of the light curve variance. Also the mean count rate of light curve would change after the exponential transformation, and a correction factor needs to be multiplied to the non-linear light curve.

\subsection{Energy dependence of the PDS}\label{sec_pdsen}

\begin{figure}
\includegraphics[width=84mm]{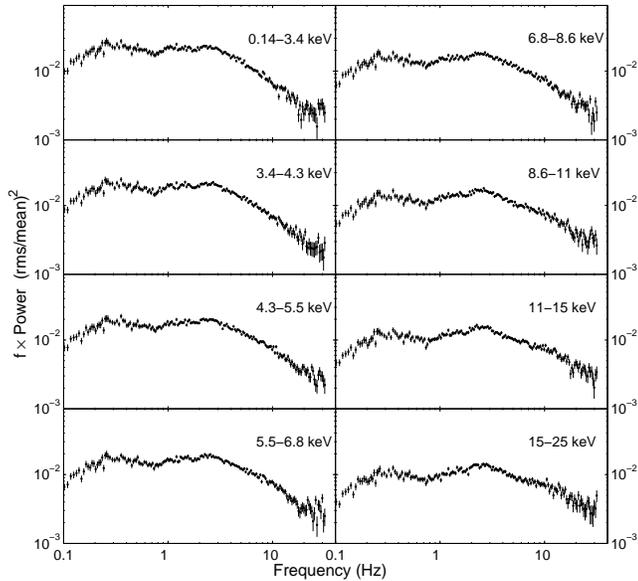}
  \caption{ The PDS in 8 energy bins, with the Poisson noise subtracted.
\label{fig_pdsfrq}}
\end{figure}

\begin{figure}
\includegraphics[width=84mm]{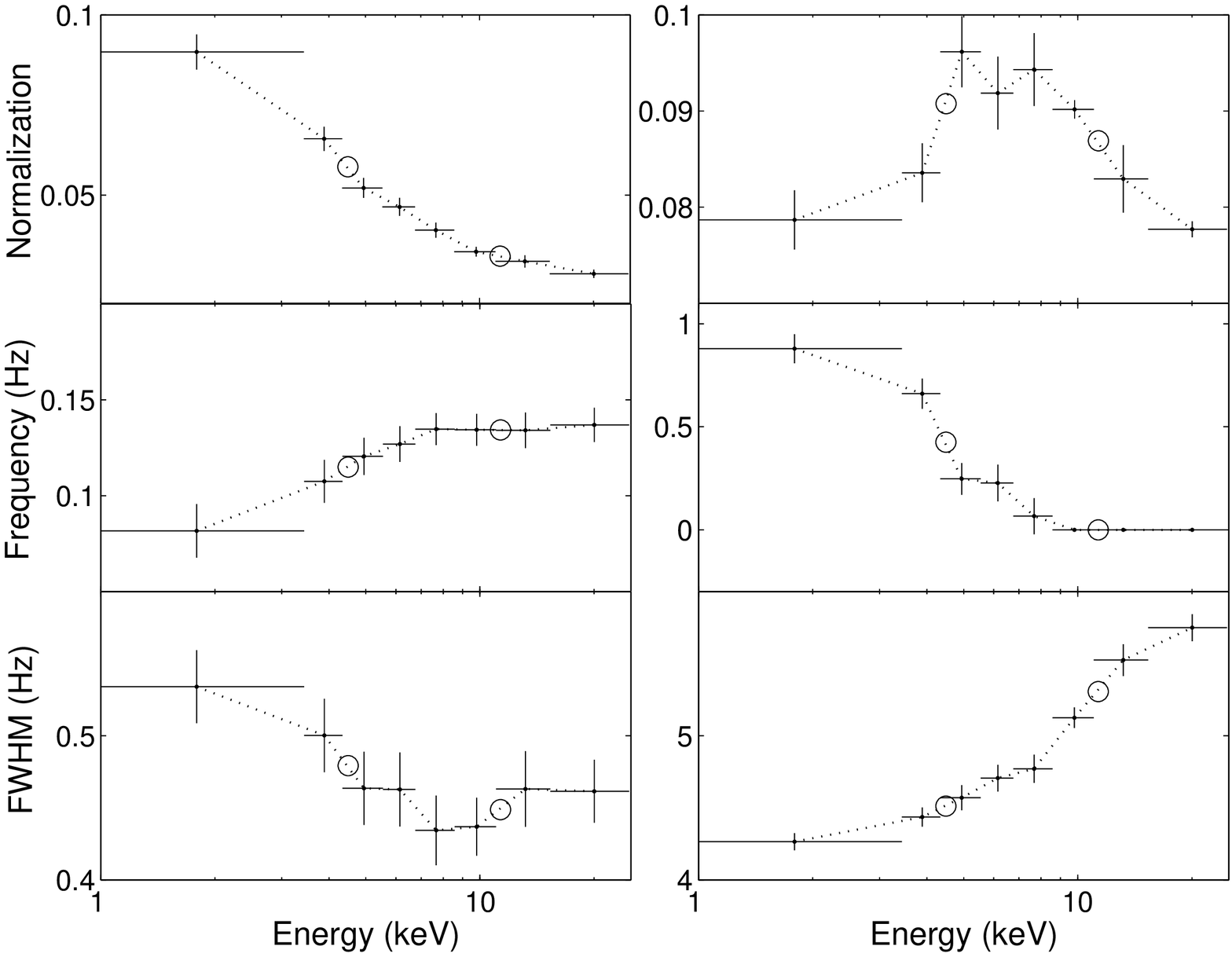}
  \caption{The best-fitting parameters of the two-Lorentzian model for the PDS at different energies. The left and right columns correspond to the two Lorentzian components. The parameters are (from top to bottom) normalization, centroid frequency and FWHM.  The circles are the interpolation values between energy bins (see section~\ref{sec_extint}).
\label{fig_pdsen}}
\end{figure}

We extracted an average PDS from each of the eight energy bins, covering the frequency range 0.008--32~Hz (Fig.~\ref{fig_pdsfrq}). No narrow QPOs are seen. A simple model consisting of two broad Lorentzians was used for the fit. The goodness of the fit is reasonably good, with all reduced $\chi^2$ smaller than 2.2 (obtaining a formal reduced $\chi^2$ of the order of unity is difficult for these high-signal PDS). Adopting a more complex model \citep[e.g. ][]{Bel97} the goodness of fit would be
improved but the fit parameters would be poorly constrained. There are
three free parameters for each Lorentzian: the normalization (the
square of the integrated fractional rms), the centroid frequency and FWHM. The evolution of the PDS shape with energy can be well described by the energy dependence of the best-fitting
parameters, which is shown in Fig.~\ref{fig_pdsen}. Concerning the second
Lorentzian, for the last three energy bins the best-fitting centroid
frequency decreases to zero, the lower bound of this free parameter.
For those three bins, uncertainties were not plotted.

\subsection{Energy dependence of the time lag spectrum}\label{sec_lagen}
The time lag of the light curve in each energy bin relative to that
of the lowest energy bin (0.14--3.4~keV) was calculated from the cross-spectra between 0.06 and 30~Hz. Positive lags here correspond to the hard time series lagging the soft.
The lags were logarithmically rebinned in frequency in order to reduce noise. Since their calculations involve the splitting of the data into two energy bands, compared with the PDS the measurement of time lags is more sensitive to counting noise. \citet{Now99} estimated the expected noise level for the time lag measurements and concluded that for frequencies below $\sim0.1$~Hz and above $\sim30$~Hz lags cannot be measured because
of noise limitations. As the frequency approaches $\sim0.1$~Hz or
$\sim30$~Hz, the lags tend to zero due to the effect of
noise. When sampling fluctuations become comparable to the intrinsic lags, they scatter around zero and exhibit negative values.

\begin{figure*}
\includegraphics[width=0.7\textwidth]{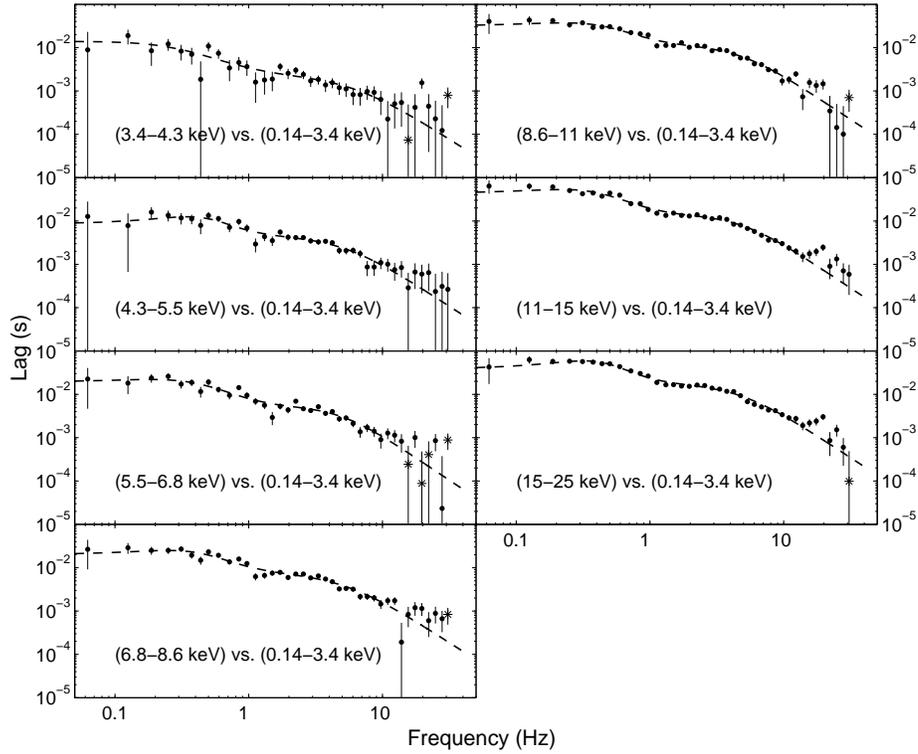}
  \caption{Time lag spectra for various energy bins versus the lowest energy bin (0.14--3.4 keV). Dots represent the positive lags (hard lagging the soft), and asterisks represent negative lags (soft lagging the hard). The dashed lines are the best-fitting two-Lorentzian models for the positive lags.
\label{fig_lagfrq}}
\end{figure*}

\begin{figure}
\includegraphics[width=84mm]{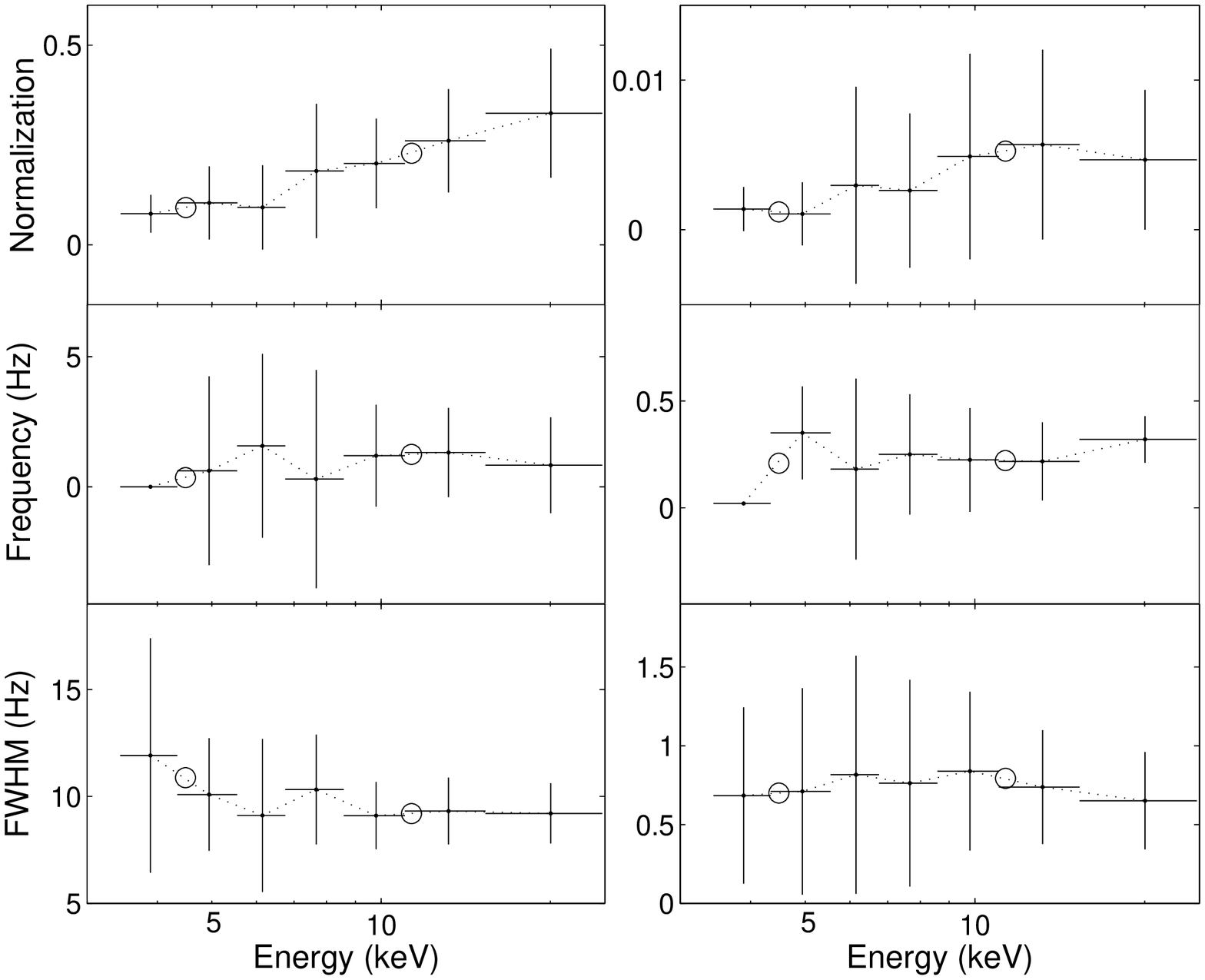}
  \caption{The best-fitting parameters of the two-Lorentzian model for time lag spectra at different energies. The left and right columns correspond to the two Lorentzian components. The parameters are (from top to bottom) normalization, centroid frequency and FWHM. The circles are the interpolation values between energy bins (see section~\ref{sec_extint}). \label{fig_lagen}}
\end{figure}

We adopted the same strategy as for the PDS to quantitatively describe
the energy dependence of lag spectrum in a uniform way.
\citet{Now99} showed that the time lags
approximately show a power law dependence upon frequency ($\propto
f^{-0.7}$). We found significant deviations from a simple power law model.
The time lag spectra show a two-humped shape similar to that
published in previous studies \citep[e.g., ][]{Miya92,Cui97,Now99}. We used a two-Lorentzian model to fit the time lag spectra. The time lag
spectra with the best-fitting two-Lorentzian models are shown in
Fig.~\ref{fig_lagfrq}. Because the negative lags have small values and appear in the frequency range where the noise level dominates, they are not expected to be intrinsic. The negative lags were therefore excluded from the fit. The evolutions of the best-fitting
parameters with energy are shown in Fig.~\ref{fig_lagen}.

\subsection{Coherence function}\label{sec_coh}
\begin{figure}
\includegraphics[width=84mm]{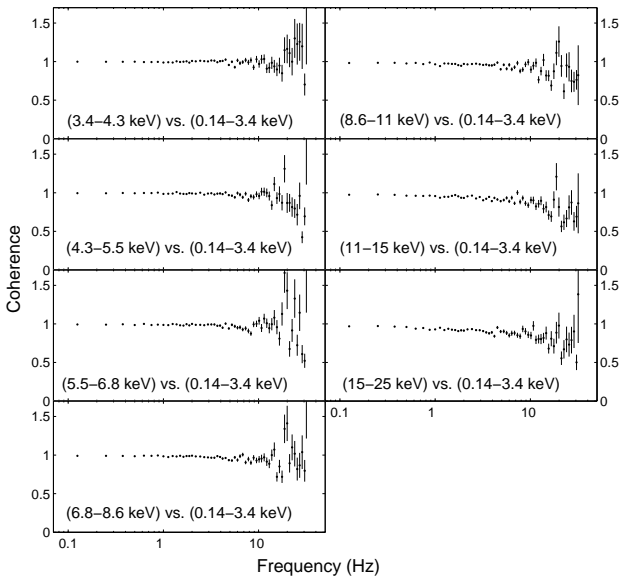}
  \caption{ The coherence function for various energy bins versus the lowest energy bin (0.14--3.4 keV).
  \label{fig_coh}}
\end{figure}

The coherence function is a Fourier frequency-dependent measure of the linear correlation between time series measured simultaneously in two energy bands\citep{VN97}. Our simulation does not include any incoherent variability, i.e. variations in one energy band that are not correlated with variations in other bands. In other words the algorithm contains an underlying assumption of a single emission component in different energy bands with a single delay at a given frequency and unity coherence. We calculated the coherence function of Cyg~X-1 data with correction for counting noise, following the recipe presented in \citet{VN97}. The results are shown in Fig.~\ref{fig_coh}, which demonstrate a remarkably high coherence (close to unity) over a wide frequency range and consistent with previous coherence study of Cyg~X-1 \citep[e.g.][]{VN97,Cui97,Now99}. Therefore, we can say that below $\sim$10 Hz, the flux in each energy band can be regarded as originating from one single coherent component, whose intrinsic phase delay is indicated by the lag spectrum. The coherence becomes slightly lower at higher frequencies, which may indicate the presence of incoherent components. We do not attempt to add them into the simulation because it is a laborious and model-dependent process and beyond the scope of this work.

\subsection{Phase}\label{sec_phas}
In order to reconstruct the time series from the PDS and time lag
spectra, a prior phase distribution has to be assumed, since the PDS
does not contain phase information. We first
analyzed the real data in order to derive a reasonable phase
distribution. We split the 0.015625-s binned light curve
into 694 segments, each with a length of 1024 points (16 seconds). For each segment, we produced a Fourier transform, leading to 694 values of
the phase angle for each frequency between 0.0625~Hz and 32~Hz.
Obviously, the phases between separate segments are comparable only after considering the additional phase shift caused by the time-delay between their start time. If $\varphi_{j,i}$ is the phase angle at frequency $f_{j}$ of the {\it
i}th segment, $t_{0,i}$ and $t_{0,1}$ are the start times of the
{\it i}th segment and the 1st segment respectively, the ``absolute''
phase at this frequency for the {\it i}th segment can be calculated
as
 \[
 \varphi_{j,i}'=\varphi_{j,i}+2\pi f_{j}(t_{0,i}-t_{0,1}).
  \]

\begin{figure*}
\includegraphics[width=0.7\textwidth]{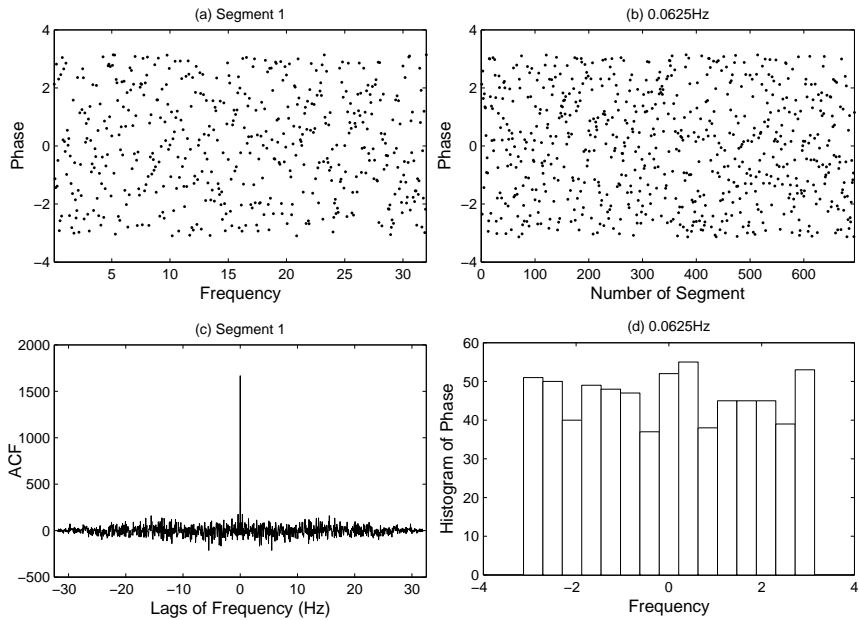}
  \caption{(a) The phase at different frequencies
for the 1st segment; (b) the phase at 0.0625~Hz for 694 segments;
(c) the auto correlation function of phase at different frequencies
for the 1st segment; (d) the histogram of phase at 0.0625~Hz.
 \label{fig_phas}}
\end{figure*}

In Fig.~\ref{fig_phas}, we plot the phase at different frequencies
for the first segment (panel {\it a}), the phase at a certain
frequency for different segments (panel {\it b}) and its histogram
of occurrence (panel {\it d}). Moreover, we studied the auto
correlation function of the phase at different frequencies (shown in
panel {\it c}). All these results clearly show that the phase
follows a uniform distribution between $-\pi$ and $\pi$£¬and the
phases at different frequencies are random and independent.
Therefore, in our synthesis algorithm we generate uniformly
distributed random numbers in the interval $(-\pi,\pi]$ as phase
angles for the Fourier transform.

\section{RESULTS}\label{sec_result}
\subsection{The algorithm}\label{sec_alg}
Having obtained the energy-resolved PDS (with Poisson noise
subtracted) and the time lag spectra for our eight energy
bins, and the average energy spectrum, we followed the following
procedure to generate a synthetic light curve:

\begin{itemize}

\item {\it Step 1}: for the lowest energy bin (1-bin), we generated uniformly-distributed
random numbers between $(-\pi,\pi]$ to be used as phase angle
$\varphi(f_i)$ at Fourier frequency $f_i(f_i\geq0)$. In order to obtain real
values for the time series, we chose the phase for the negative
frequencies as $\varphi(-f_i)=-\varphi(f_i)$. For the energy
bins 2 through 8, the phases were reconstructed from the 1-bin values according to
the measured phase lag spectra which provide the phase shift
relative to the lowest energy bin at each frequency.

\item {\it Step 2}: the amplitude of the Fourier transform at each frequency
was obtained from the PDS. In order to account for the effects of the
exponential transformation to the variance of light curve, the PDS
should first be renormalized in order to obtain the desired variance. For a PDS $P(f)$ in units of (rms/mean)$^2$~Hz$^{-1}$ and
with frequency bin size $\Delta f$, the desired fractional rms is
$R^2=\Sigma P(f)\Delta f$. The PDS must be multiplied by a
factor of $\log(R^2+1)/R^2$ \citep[for details see][]{Utt05}. The square root of the renormalized PDS
is the amplitude of the Fourier transform $A(f_i)$. The series need
also be expanded to negative frequencies with $A(-f_i)=A(f_i)$.

\item {\it Step 3}: for each energy band, we calculated the inverse
Fourier transform of $A(f_i)\exp(j\varphi(f_i))$ (where $j$ is the
imaginary unit) to obtain the linear time series $l(t)$, and then
calculated its exponential. In order to ensure that the simulated
light curve has the desired mean count rate $C(E)$ measured in the
average energy spectra, a factor of $C(E)\log(R^2+1)/R^2$ needs to
be multiplied to $\exp[l(t)]$.

\item {\it Step 4}: the time series obtained with the previous steps
was stored as one light-curve segment. Steps of 1--3 were then
repeated to produce multiple segments.

\end{itemize}

\subsection{Test of the Simulation}\label{sec_test}

\begin{figure*}
\includegraphics[width=0.7\textwidth]{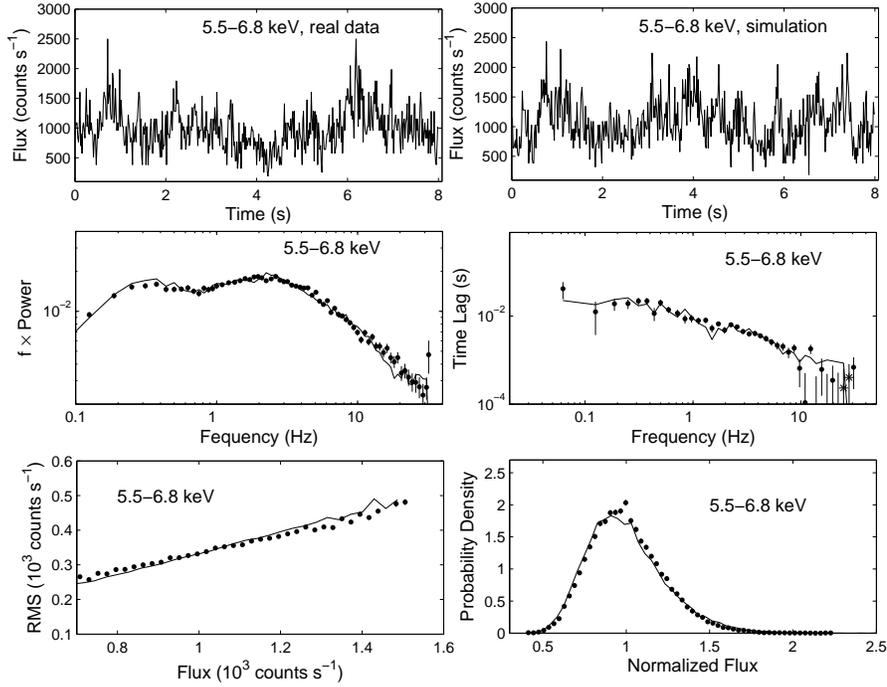}
  \caption{Comparison of the simulation with the
algorithm described in section~\ref{sec_alg} (with additional
Poisson noise) and the real data (ObsID 10238-01-05-000) for the energy bin 5.5--6.8~keV. Top: one segment (1024 points) of light curve of the real
data (left) and the simulation (right). Middle: the PDS (left) and the
time lag spectra (right) of the simulation (dot for positive
values and asterisk for negative values) and the real data (solid
line). Bottom: the rms-flux relation (left) and flux distribution
(right) of the simulation (dot) and the real data (solid line). The error
bars of the simulation data are plotted.
\label{fig_sim}}
\end{figure*}

To check whether the simulated light curve replicates all the
observed properties of original real data, we compared their PDS,
time lag spectra, rms-flux relation and lognormal flux distribution.
The comparison relative to the energy bin 5.5-6.8~keV is shown in
Fig.~\ref{fig_sim} as an example. The simulated light curve of
course cannot have exactly the same evolution as the real one, as a
random-number input is involved,
but they appear to be similar in the amplitude and time-scale of
variance. The PDS, the time lag spectrum, the rms-flux relation and
the flux distribution are consistent with those from the real data,
showing that our simulation reproduces accurately the intrinsic
properties of the real data. In other words, our algorithm can
synthesize data whose statistical properties are indistinguishable
from those observed from Cyg~X-1.

\subsection{Poisson Noise Subtraction}\label{sec_subnoise}
The expected influence of Poisson fluctuations in the time
series is represented as a white noise component in the PDS. Since
it is independent of the source signal (apart from dead-time
effects), the Poisson noise can be considered as a ``background''
component in the PDS, against which we try to observe other features
caused by the intrinsic variability of X-ray source. If the light
curve is a series of contiguous time bins, the expected Poisson
noise level is simply 2 for the ``Leahy'' normalization
\citep{Lea83}. In this work the PDS is normalized in units of (rms/mean)$^2$~Hz$^{-1}$
\citep{BH90}, and the
expected Poisson noise level in the PDS is given by
\[
P_{\rm noise}=\frac{2(C+B)}{C^2}
 \]
where $C$ and $B$ are the mean source count rate and background
count rate, respectively. See \citet{Vau03} for more details about
the different normalizations of PDS and the corresponding Poisson noise
levels. Notice that the shape and level of the Poisson noise contribution
to the PDS are modified by dead-time effects \citep[for the {\it
RXTE}/PCA, see][]{Zha95}.

Therefore, we can easily subtract the Poisson noise level from the PDS,
and obtain the ``clean'' light curve without Poisson noise with our
algorithm. In other words, our synthetic algorithm can be used as a
filter of Poisson noise. We can check this by adding Poisson
fluctuations to an initial simulated ``clean'' light curve, then
filter the ``dirty'' light curve with our algorithm to see whether
the filtered light curve resembles the initial one or not. Because
the algorithm can not repeat the exact shape of the light curve due
to the phase randomization, in order to make a direct comparison
between the clean light curve and filtered light curve, we have to
record the phase information of the
initial data as the input of the algorithm instead of using random
phases. We do this here because our aim is simply to check the
effect of Poisson noise filtering for the synthetic algorithm.

\begin{figure*}
\includegraphics[width=0.7\textwidth]{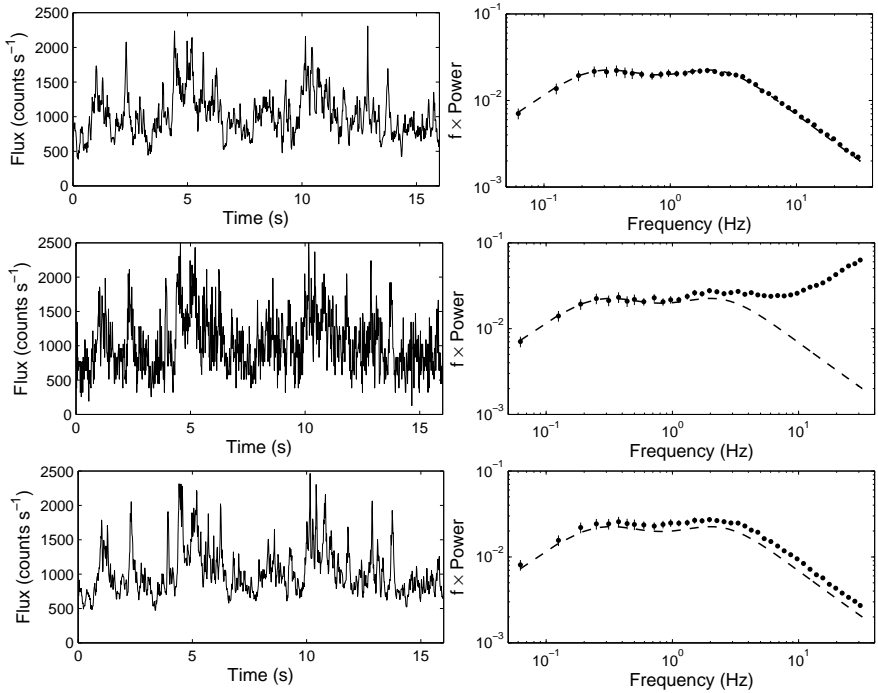}
  \caption{Top: the initial clean light curve and
its PDS. Middle: the light curve with additional Poisson
noise and its PDS. Bottom: the light curve filtered by our algorithm
and its PDS. The dashed line in the right panels is the PDS of
the initial clean light curve. The PDS are calculated from 50
segments, each with a length of 1024 points. Each of the left panels
shows only one segment of the corresponding light curve. \label{fig_filter}}
\end{figure*}

The results are shown in Fig.~\ref{fig_filter}. We can see how the
Poisson noise is removed in the filtered light curve. A quantitive
measurement is the standard deviation, which is 320, 420 and 340
for the clean, dirty and filtered light curves respectively. The
excess variation of the filtered light curve at high frequencies
is very probably caused by the distortion introduced by the exponential
transform, which tends to exaggerate the positive variation, e.g.
the amplitudes of the flares. The conclusion is that the synthetic
light curve created by our algorithm can be considered essentially
free of Poisson noise.

It is worthy to point out that if the phases were known, we could
reconstruct the time series strictly by inverse Fourier transform,
because this time-to-frequency transition is completely reversible.
The exponential transform would be not be necessary.
We therefore face an interesting problem: if the phase is known, the exponential is redundant for reconstructing the initial time series, which is completely defined by the fourier spectrum; however if we know nothing about the phases and assume them to be random, in order to reproduce a time series satisfactorily we have to apply the exponential transformation. We will discuss this problem in section~\ref{sec_posnois}.

\begin{figure}
\includegraphics[width=84mm]{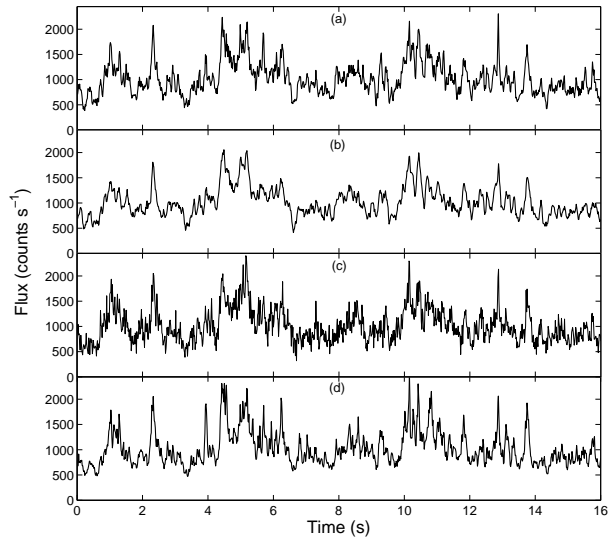}
  \caption{ (a): the initial clean light curve (one segment, the same as in the left top panel of Fig.~\ref{fig_filter}). This is used as the input for both Wiener filtering and our algorithm, after adding Poissonian noise (see text). (b): the Wiener filtering solution with the averaged PDS over many segments. (c): the Wiener filtering solution with the PDS of the single segment. (d): the solution of our algorithm (the same as in the left bottom panel of Fig.~\ref{fig_filter}). \label{fig_wnr}}
\end{figure}

The above process of subtracting the Poisson noise is similar to Wiener filtering. The Wiener filter is the optimal filter in the least-square sense for the removal of noise from a time domain signal. The Wiener filter is designed in the Fourier domain and can be expressed as \citep{Pre92}:
 \[
\Phi (f)=\frac{|S(f)|^2}{|S(f)|^2+|N(f)|^2}
  \]
in which $S$ and $N$ are the Fourier transforms of the intrinsic signal and the noise, respectively. The denominator $|S(f)|^2+|N(f)|^2$ is proportional to the PDS of the measured light curve (under the assumption that signal and noise are statistically independent). The filter can be constructed if the true form of the intrinsic power $|S(f)|^2$ is known or can be estimated well. Our algorithm and the Wiener filter share some common ideas --- to separate noise and signal in the frequency domain, which can not be done in the time domain. We fixed the noise power to 2 and applied a Wiener filter to the same data in Fig.~\ref{fig_filter}. The difference between the Wiener filtering solution and ours is shown in Fig.~\ref{fig_wnr}. The filtering effect are generally similar for the two methods. If the PDS used in the Wiener filtering and our algorithm are those averaged over many segments (panel {\it b} and {\it d} in Fig.~\ref{fig_wnr}), our algorithm appears to preserve more short time-scale fluctuations, which makes it more similar to the original one. If the PDS of the single segment shown in the figure is used in designing the Wiener filter (panel {\it c} in Fig.~\ref{fig_wnr}), its solution would be much more noisy than the original light curve. From this perspective, our algorithm seems to provide a more stable filter as it is able to subtract the noise and at the same time to avoid wiping off too much rapid variability. This is probably due to the exponential transformation that can restore the fluctuation amplitude to a certain extent. The price of the exponential transformation is that the exact shape of light-curve is slightly distorted.

\subsection{Dynamic Energy Spectra}\label{sec_des}

If the photon count in one bin of the energy spectrum is $N$, the
relative standard deviation expected from the Poisson distribution
is $1/\sqrt{N}$. When we want to obtain the average energy
spectra with high statistical significance, we need a sufficiently
long exposure (hundreds to thousands of seconds) to accumulate
enough photons. It is important to see how the energy
spectra evolve on short time-scales which, however, in this way is not possible. The energy spectra on short time-scales, which we name dynamic energy spectra, can be obtained by aligning the light curves of different energy bins in time and obtaining the counts in every time bin as a function of energy. The problem of this analysis from the real data is that the Poisson fluctuation is severe in this case due to the small number of counts. The simulated light curve obtained with our synthetic
algorithm, as presented in the last section, does not include Poisson noise. The dynamic energy spectrum produced from the simulated light curves is therefore ``cleaner'' and can reveal the underlying properties
otherwise hidden by noise.

\begin{figure*}
\includegraphics[width=0.8\textwidth]{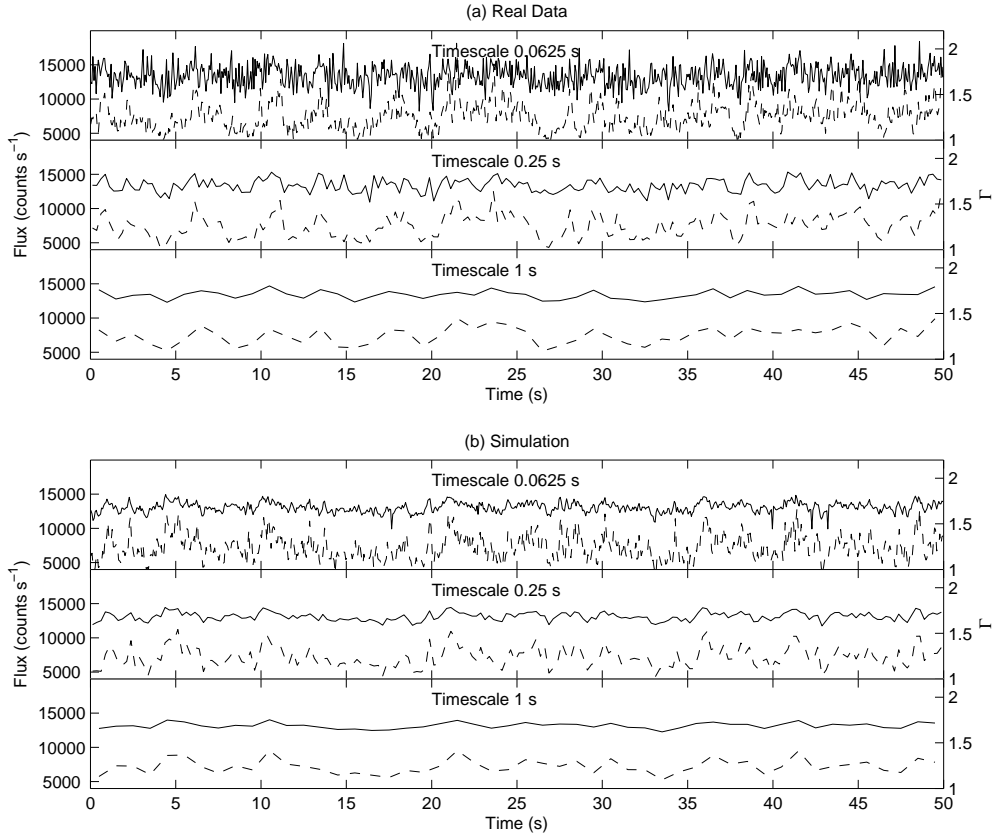}
  \caption{The 0.14--25~keV light curve (dashed
line) and the corresponding power law photon index $\Gamma$
evolution (solid line) for (a) real data and (b) simulation (without
Poisson noise), with time bin sizes of 0.0625~s, 0.25~s and 1~s from
top to bottom. \label{fig_fluxpl}}
\end{figure*}

\begin{figure}
\includegraphics[width=84mm]{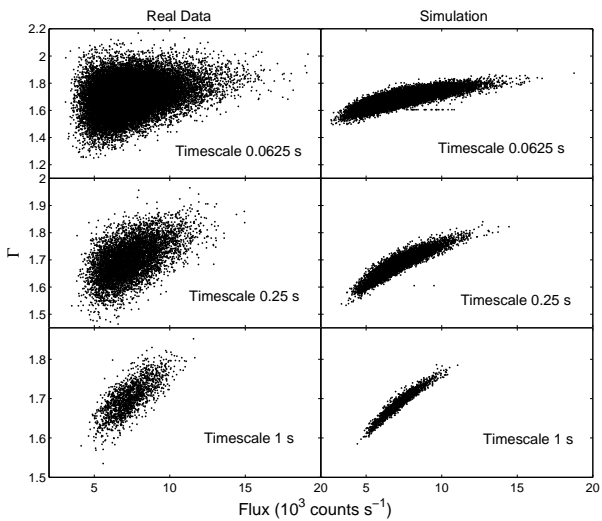}
  \caption{The correlation between power law photon
index $\Gamma$ and flux at three time-scales (or time bin sizes)
of 0.0625~s, 0.25~s and 1~s for real data (left column) and
simulation (right column). \label{fig_corr}}
\end{figure}

\begin{figure}
\includegraphics[width=84mm]{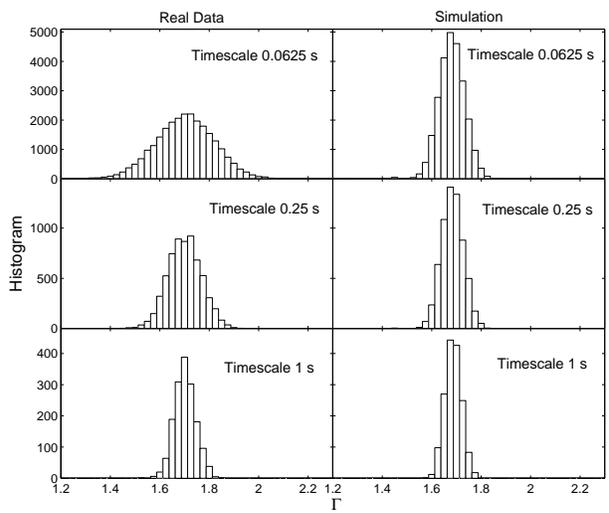}
  \caption{The histogram of power law photon index
$\Gamma$ at three time-scales (or time bin sizes) of 0.0625~s, 0.25~s
and 1~s for real data (left column) and simulation (right column).
\label{fig_hispl}}
\end{figure}

The dynamic energy spectra with 8 energy bins were calculated for
both real data and simulation at three time-scales (or time bin
sizes), 0.0625~s, 0.25~s and 1~s, and then fitted with XSPEC using
the PCA detector response matrix. Notice that the synthetic data,
having reproduced the background-subtracted energy spectrum, are
also background free. The lowest energy bin (0.14-3.4~keV) is
excluded due to the uncertainty in the PCA calibration below 3 keV
and a simple power law is fitted to each of the dynamic energy
spectra. There are in total 25600, 6400 and 1600 dynamic energy
spectra that were fitted for the three time-scales, respectively. The
power law photon index $\Gamma$ is the parameter that we studied to
reflect the basic shape of dynamic energy spectra. For a comparison
of real data and simulation, we plot the time evolution of $\Gamma$
as well as the flux covering the whole energy band (0.14--25~keV)
(Fig.~\ref{fig_fluxpl}), the correlation between $\Gamma$ and flux
(Fig.~\ref{fig_corr}) and the histogram of $\Gamma$
(Fig.~\ref{fig_hispl}). The noise-free reconstructed data providesa ``cleaner'' view of the $\Gamma$ distribution and the $\Gamma$-flux correlation. One possible reason for the improved correlation is that the simulation does not include any incoherent variations that may weaken the correlation for the real data. However, we have shown that the coherence in the real data is very close to unity for most of the frequencies considered here and therefore this possibility can be excluded.

Combining the above results we can conclude that:
\begin{enumerate}
\item The correlation between $\Gamma$ and flux is somewhat
higher for the simulation than for the real data.
\item The
distribution of $\Gamma$ is narrower for the simulation than for the
real data.
\item The above differences between simulation and real data
tend to increase at shorter time-scales, i.e. for lower photon count numbers.
\end{enumerate}

The correlation between $\Gamma$ and flux is consistent
with the previous results that the hardness ratio anti-correlates
with the X-ray flux \citep[e.g.][]{Cui02,LL04,Wil06} or that the
photon index correlates similarly with the flux on time-scale of days
\citep[e.g.][]{Zdz02,Pot03,GZ03}. The significantly better
correlation for simulated data (Fig.~\ref{fig_corr}) shows that
our algorithm enables us to significantly reduce the effects of Poisson noise and study the intrinsic spectral evolution at short time-scales. The Poisson noise also broadens significantly the distribution of
$\Gamma$ (Fig.~\ref{fig_hispl}). However, we need to be cautious
to claim that the broadening of the $\Gamma$ distribution from the
simulation (as shown in the right column in the Fig.~\ref{fig_hispl}) is completely caused by the intrinsic short-time-scale spectral fluctuation. It might also be introduced by our algorithm, which we investigate next.

\begin{figure*}
\includegraphics[width=0.7\textwidth]{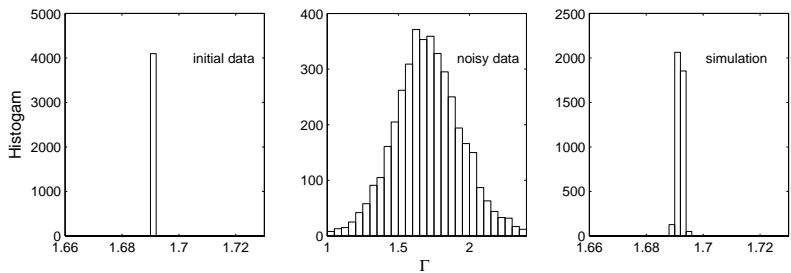}
  \caption{The histograms of power low photon index $\Gamma$ at 0.0625~s for data with $\Gamma$ distributed as a $\delta$ function at 1.69 (left), the same data with added Poisson noise (middle) and the reconstructed data with our synthetic algorithm (right).
\label{fig_disphidx}}
\end{figure*}

In order to test the effects of the algorithm on the recovery of
$\Gamma$ values, we produced a new synthetic dataset by using all
information described above with the exception of the $\Gamma$
values, which were fixed at the single value of 1.69. In other
words, we produced a set of light curves with null time/phase lags,
which were therefore identical except for their normalizations. We added
Poisson noise and studied the output $\Gamma$ distribution after
applying our algorithm. The histograms of $\Gamma$ for the zero-lag
synthetic data, the values obtained through direct spectral fitting
and those from our reconstruction are plotted in Fig.~\ref{fig_disphidx}. From our test, we can derive that the systematic broadening of the $\Gamma$ distribution introduced by our procedure is very limited (less than 0.004 or 0.24 per cent) (notice the horizon scale of the right panel in Fig.~\ref{fig_disphidx}).

The observed $\Gamma$ distribution shown in the left column of Fig.~\ref{fig_hispl} is the result of both the intrinsic distribution in $\Gamma$ and the broadening due to noise. Thus, the standard deviation $\sigma_{\Gamma}$ of the intrinsic spread in $\Gamma$ can be quantitatively estimated as:
 \[
\sigma_{\Gamma}=\sqrt{\sigma_{\rm data}^2-\langle e^2 \rangle }
  \]
where $\sigma_{\rm data}^2$ is the variance of the $\Gamma$ distribution observed from real data (left column of Fig.~\ref{fig_hispl}), and $\langle e^2 \rangle$ is the mean square error on $\Gamma$ obtained from the spectral fitting analysis. For the shortest time-scale 0.0625~s, we have obtained 25600 values of $\Gamma$ from fitting the dynamic energy spectra, obtaining $\sigma_{\rm data}^2=0.0141$, $\langle e^2 \rangle=0.0113$. $\sigma_{\rm data}^2$ is approximately equal to $\langle e^2 \rangle$, which proves that the spread in $\Gamma$ at short time-scale from the real data is mostly due to the Poisson noise. With the above equation we obtain $\sigma_{\Gamma}=0.0529$, close to the standard deviation of 0.0517 calculated from the simulation data (the top right panel in Fig.~\ref{fig_hispl}). For the longer time-scales, the two values are also found to be comparable. Again this fact supports that our algorithm is capable of filtering the Poisson noise and reveal the intrinsic distribution of $\Gamma$. It is interesting to notice that \citet{Kot01} assumed the small variations of the power law index in their time-lag model. Our study on the intrinsic $\Gamma$ distribution thereby gives an evidence for their assumption.

We can therefore draw three conclusions:
\begin{enumerate}
\item  the broadening of $\Gamma$ distribution introduced
by our algorithm can be neglected and the histograms on the right
panel of Fig.~\ref{fig_hispl} reflect the intrinsic spectral
variation of the source on these time-scales;

\item our synthetic method is indeed powerful in filtering Poisson
noise and recovering underlying statistical properties of the
source data which are masked by Poisson noise;

\item the $\Gamma$ distribution obtained through short-time
spectral fitting of the data is completely dominated by the effects of Poisson noise and cannot be used to ascertain the real
distribution.
\end{enumerate}

A different approach was followed studied by \citet{Rev99} through Fourier-resolved spectra, which give the energy-dependent variability amplitude in a certain frequency range. The frequency dependent spectral variability revealed by the Fourier-resolved spectra, however, cannot be immediately linked with the variation of spectral indices studied here. First, the time-scale in our work refers to the time bin size other than the reciprocal of frequency. The variations sampled on short time bins come from both low-frequency and high-frequency variabilities presented in PDS. However, the power density or variability calculated by binning the time series is comparable to that on the corresponding frequency \citep[see][and refernces therein]{wu09}, and the time-scale can be taken in practice as the time bin size. What is more important is that although Fourier-resolved spectra have a form similar to energy spectra, they have a totally different physical interpretation. Therefore, the comparison of their spectral indices is not useful. For example, the Fourier-resolved spectrum was found to be harder at higher frequencies, which only suggests that the hard X-ray radiation component exhibits larger variation amplitude compared with the soft radiation as frequency increases.
The only possible connection we can seek between this phenomenon and our study on the energy spectral indices is that the variations of $\Gamma$ on short time-scale are probably mainly due to the hard spectral component.

\subsection{Extrapolation and Interpolation}\label{sec_extint}
The synthetic algorithm can be used to produce the data with better
time and energy resolution than the real data, after some additional
hypotheses are made. A two-Lorentzian model has been used to
describe the PDS and lag spectra (see section~\ref{sec_pdsen} and
\ref{sec_lagen}). If we extrapolate the model frequency beyond the
Nyquist frequency of the original data, we are able to derive a
synthetic light curve with higher time resolution. Moreover, we have
derived the energy dependence of the best-fitting parameters for the
PDS and lag spectra. By interpolating these functions, we can obtain
the PDS and lag spectra (and therefore produce synthetic data)
on an energy grid finer than the initial energy resolution. The
hypothesis on which the extrapolation and interpolation are based is
of course that PDS and time lag evolve smoothly in frequency and
energy and that they can be extrapolated from the observed values.

\begin{figure*}
\includegraphics[width=0.7\textwidth]{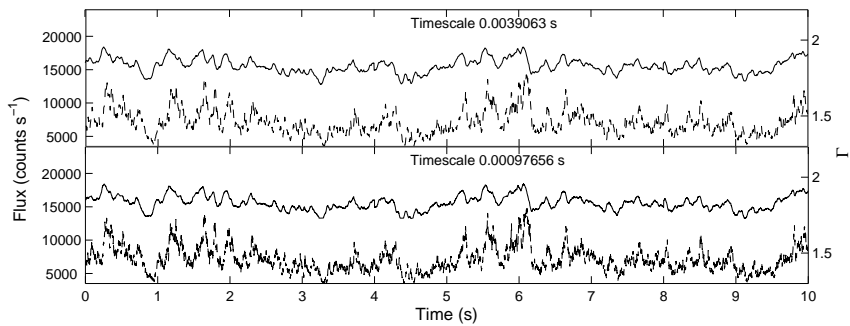}
  \caption{The time evolution of power law photon index
$\Gamma$ (solid line) and 0.14--25~keV net flux (dashed line) derived
from the simulation based on the extrapolation of PDS and time lag
spectra to higher frequency. The time-scales of 0.004~s (upper) and
0.001~s (lower) are shorter than the time resolution of the original
data. \label{fig_ext}}
\end{figure*}

\begin{figure}
\includegraphics[width=84mm]{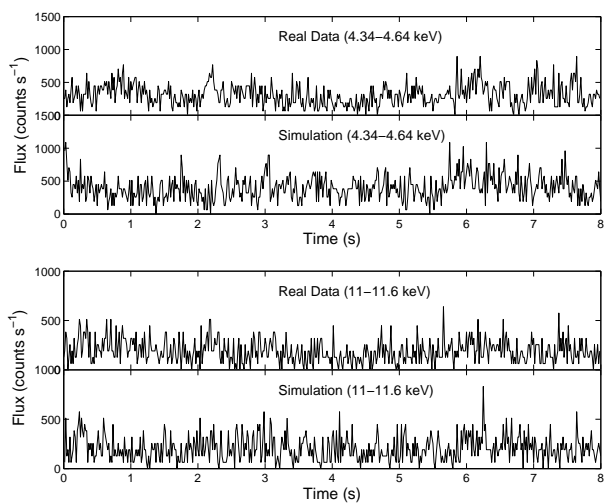}
  \caption{The real and simulated light
curves in two narrow energy bins of 4.34--4.64~keV (upper panel) and
11--11.6~keV (lower panel). The simulation is based on the interpolation
of the best-fitting parameters for PDS and time lag spectra between
energy bins. \label{fig_int}}
\end{figure}

The results of the extrapolation to higher frequency and of the
interpolation between energies are shown in Figs.~\ref{fig_ext} and
\ref{fig_int} respectively. The time resolutions in the two panels
of Fig.~\ref{fig_ext} are 0.004 and 0.001~s, smaller than the time
resolution of 0.015625~s for the real data. The dynamic energy
spectra were also studied on these time-scales and the resulting
power law photon indices are plotted in the figure. Poisson noise is
{\it not} added to the light curve and an explicit positive
correlation between photon index and flux can be observed. The
best-fitting parameters of PDS and lag spectra were interpolated to
two additional narrow energy bins, 4.34--4.64~keV and 11--11.6~keV
(see open circles in Fig.~\ref{fig_pdsen} and Fig.~\ref{fig_lagen}). As
mentioned, the original data have 64 energy bins, which we rebinned
into 8 coarse energy bins as the input of our algorithm. This
allowed us to compare the simulation and the real data in these two
narrow energy bins, as shown in Fig.~\ref{fig_int}.

\subsection{Phase and Noise}\label{sec_posnois}

In section~\ref{sec_subnoise}, we wondered whether an exponentiation is necessary after the inverse Fourier transform to produce a light curve similar to the real one when random phases are assumed. This is peculiar since we can strictly recover the time series with the inverse Fourier transform alone, if we know the phases. It is therefore logical to deduce that the phases of the real data cannot be random. The exponential transformation is simply a compensation for the incorrect assumption of random phases. The effect of the exponential transformation in the time domain should be represented in the Fourier domain as a modification of phase, since there is practically no effect on the PDS shape. The fact that the phase cannot be totally random and independent can also be revealed by higher order variability properties, e.g. the bicoherence, a higher order statistics measuring the degree of coupling between variations on different time-scales. A non-zero bicoherence indicates that there exist correlations between the phases at different frequencies within a single energy band \citep{MC02,Utt05}. Hence the phases cannot be independent, nor can they be strictly uncorrelated. From Fig.~\ref{fig_phas} the phases appear to be uniformly distributed over the range $(-\pi,\pi]$ and a correlation between them is not apparent. A higher order statistical test such as bicoherence would probably show these effects.
The assumption about the phase distribution presented in section~\ref{sec_subnoise} can be considered as a lower-order approximation, and is appropriate for the practical purpose of simulating the linear time series which will later be  exponentially transformed.

\begin{figure*}
\includegraphics[width=0.7\textwidth]{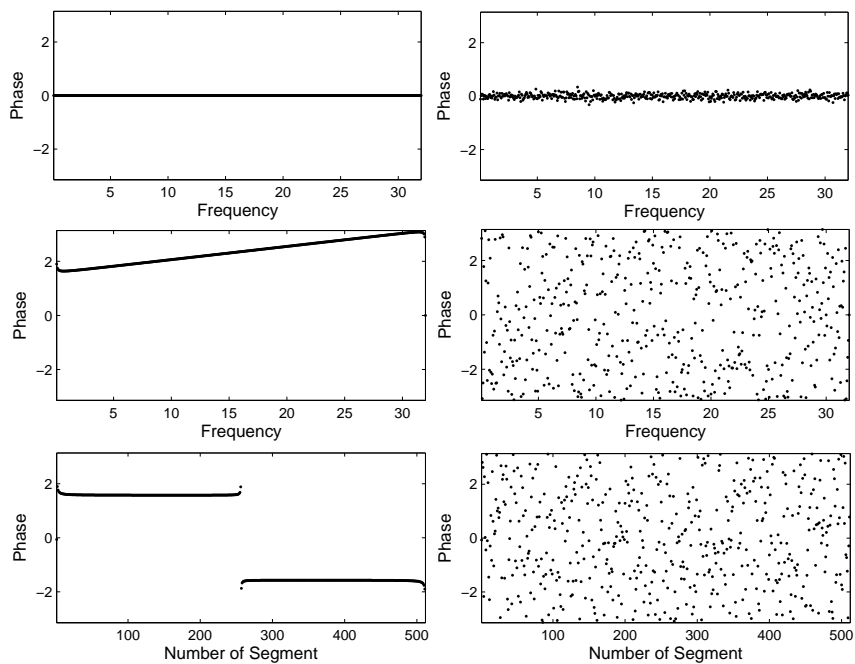}
  \caption{Top: the phases of the initial light curve with
(left) and without (right) additional Poisson noise. Middle:
the phases at different frequencies for a single segment after
arbitrarily choosing the starting point and segment (left) and the case
with additional Poisson noise (right). Bottom: the phase for
different segments for a fixed frequency
after arbitrarily choosing the starting point and segment (left) and
the case with additional Poisson noise (right).
\label{fig_phasnois}}
\end{figure*}

Moreover, it is possible that the procedure of segmenting the data conceals the underlying phase.
In order to investigate this possibility, we performed this experiment (see Fig.~\ref{fig_phasnois}). With an arbitrary PDS and zero lag between energies, we synthesized an initial light curve. Its time lag is of course zero at all frequencies. If we add Poisson noise to the light curve, the phases (calculated from the FFT) would be scattered around zero. In practice the start time of the observation is arbitrary and the long light curve is split into short segments to calculate the phase. We chose an arbitrary starting point and calculated the FFT in 1024s-long segments. The
phases at different frequencies for a single segment, and the phases
at different segments for a single frequency apparently deviate from
zero. If we again add Poisson noise, the phase becomes totally random. Therefore, even if the intrinsic phase is not random or uncorrelated, the arbitrary selection of a starting
point and the presence of additional Poisson noise would make the detected phase appear random. The underlying phase is likely
not random, although this cannot be inferred directly by the data. Up to now
we still know little about the intrinsic distribution of the phases
and cannot propose a hypothesis more reasonable than the random
distribution. Therefore, we stick to the assumption of random phases
uniformly distributed between $(-\pi,\pi]$ throughout this work.

\section{Conclusions}\label{sec_con}

We have developed an algorithm to produce synthetic light curves from
the observed properties of Cyg~X-1. The algorithm is based on the previously established time series simulation method \citep[e.g.][]{TK95,Utt05} and improved in three aspects: a) besides the PDS, the information from additional measurements such as energy spectra and lag spectra is used as input; b) the synthetic time series is required to reproduce almost all the timing and spectral properties; c) the possible applications of the method are explored, including filtering Poisson noise and data extrapolation. It is model independent, unlike the attempts to restore the timing and spectral properties through physically interesting model and parameters \citep[e.g.][]{AU06}. The simulation do not provide information that are not already contained in the original PDS, lag spectra and etc. What we do is to allow a different view of the same data.

By combining all known information about the observed variability, a reasonable assumption on the distribution of phases, and prior knowledges about the Poisson noise power, we can obtain synthetic data which are not affected by Poisson noise. From these synthetic data, we can explore the spectral variability of the source on short time-scales, where the real data are noise-dominated. We showed that the observed distribution of spectral indices of Cyg~X-1 on short time-scales is completely dominated by Poisson effects, as even simulated data with a $\delta$ distribution in spectral indices yields the same output distribution. From the output of our algorithm, we have recovered the real underlying distribution, under a relatively small number of assumptions. Our method shows that our current data are sufficient to reproduce the observed properties with good accuracy. Future missions will yield much higher statistics and will allow to explore spectral variability at higher frequencies and with better spectral resolution.

\section*{Acknowledgments}

We thank Phil Uttley for providing the code for the method published in \citet{Utt05}. Y. X. Wu thanks T. P. Li and S. N. Zhang for useful comments. We also appreciate the anonymous referee for the very insightful suggestions, which help improving the article a lot.
This work was supported by contract PRIN-INAF 2006.

\bsp

\label{lastpage}

\end{document}